\documentclass[aps,prl,reprint,superscriptaddress,nofootinbib]{revtex4-2}

\usepackage[utf8]{inputenc}
\usepackage[T1]{fontenc}

\usepackage{amsmath,amssymb,amsfonts, mathtools}
\usepackage{physics,overpic}
\usepackage{bm}
\usepackage{graphicx}

\usepackage[
colorlinks=true,
linkcolor=blue,
citecolor=blue,
urlcolor=blue
]{hyperref}

\usepackage[dvipsnames]{xcolor}
\usepackage[normalem]{ulem}

\begin{document}

\title{Renormalization effects fade away during inflation}

\author{Christian Dur\'an Romero}
\email{chduran@ucm.es}
\affiliation{
Departamento de F\'isica Te\'orica and IPARCOS,
Universidad Complutense de Madrid,
Plaza de las Ciencias 1, 28040 Madrid, Spain
}

\author{Luis J. Garay}
\email{luisj.garay@ucm.es}
\author{Mercedes Mart\'in-Benito}
\email{m.martin.benito@ucm.es}

\affiliation{
Departamento de F\'isica Te\'orica and IPARCOS,
Universidad Complutense de Madrid,
Plaza de las Ciencias 1, 28040 Madrid, Spain
}

\author{Rita B. Neves}
\email{rita.neves@sheffield.ac.uk}
\affiliation{
School of Mathematical and Physical Sciences,
University of Sheffield,
Hicks Building, Hounsfield Road,
Sheffield S3 7RH, United Kingdom
}


\begin{abstract}
The renormalization of the primordial inflationary power spectrum has long raised the possibility that ultraviolet effects could significantly alter predictions for cosmological observables. We demonstrate that inflation dynamically suppresses the entire renormalization sector: while super-Hubble perturbations freeze after horizon crossing, renormalization contributions decay rapidly during inflation. As a consequence, the observable primordial spectrum is remarkably insensitive to renormalization ambiguities, providing strong evidence for the robustness under renormalization of standard inflationary predictions at observable scales.
\end{abstract}

\maketitle

\textit{Introduction.---}The theory of cosmological inflation has become the standard framework for describing the physics of the early Universe. Originally introduced to explain the flatness, horizon, and monopole problems of the standard Big Bang model~\cite{starobinsky1980new,Guth,linde1982new,albrecht1982cosmology}, inflation postulates a brief stage of accelerated expansion capable of stretching microscopic quantum fluctuations to cosmological scales. Remarkably, these primordial quantum fluctuations provide the seeds for the temperature anisotropies observed in the Cosmic Microwave Background (CMB) as well as for the formation of the large-scale structure of the Universe~\cite{Mukhanov:1981xt,hawking1982,Mukhanov:1987pv,guth1982fluctuations,bardeen,MFB1992,mukhanov2005physical}. At the perturbative level, scalar cosmological fluctuations are encoded in the gauge-invariant Mukhanov-Sasaki variable, whose quantization determines the primordial two-point correlation function and the associated scalar power spectrum \cite{Sasaki:1986hm,Mukhanov:1988jd}. This spectrum provides the fundamental link between quantum fluctuations generated during inflation and the observed large-scale structure of the Universe. As one of the major successes of inflationary cosmology and quantum field theory in curved spacetimes (QFTCS), its nearly scale-invariant form is confirmed by current cosmological observations with remarkable precision \cite{wmap2011,Planck2018,achúcarro2022inflationtheoryobservations}.

However, from the viewpoint of QFTCS, the coincidence limit of the two-point correlation function is ultraviolet (UV) divergent and therefore requires renormalization \cite{wald1994qft, BirrellDavies1982, parker_toms_qft_curved_spacetime}. This issue was originally emphasized by Parker \cite{parker2007amplitudeperturbationsinflation} and subsequently explored using adiabatic subtraction and related renormalization prescriptions. Remarkably, several works \cite{Agullo__2009, del_Rio_2014, del_Rio_2015} reported that renormalization could drastically modify the amplitude of the primordial spectrum, even at observational scales, potentially generating a tension between QFTCS and cosmological observations. More recently, this conclusion has been questioned and several analyses have suggested that the observable spectrum may in fact be largely insensitive to renormalization effects \cite{Markkanen_2018, silviapla, Zhang_2018, Ferreiro_2024}. Nevertheless, the physical origin and magnitude of the remaining renormalization ambiguities remain actively debated (see, e.g.,~\cite{Markkanen_2018, silviapla, negro} and references therein).

In this work we revisit the renormalization of the scalar primordial power spectrum from a different perspective. Employing the asymptotic renormalization framework \cite{Romero:2026mno}, we show that observable inflationary predictions are extraordinarily robust against physically admissible renormalization ambiguities. The key result is that inflation itself dynamically suppresses the renormalization sector: while super-Hubble perturbations freeze after horizon crossing, renormalization contributions decay rapidly as inflation proceeds. Consequently, the observable primordial spectrum becomes largely insensitive to the details of the renormalization prescription.


\textit{Primordial power spectrum.---}
In inflationary cosmology, the primordial density perturbations observed in the CMB originate from quantum fluctuations generated during the accelerated expansion of the early Universe. At linear order, scalar perturbations are conveniently described in terms of the comoving curvature perturbation $\mathcal R(\eta,\mathbf{x})$, related to the Mukhanov-Sasaki variable $u(\eta,\mathbf{x})$ through $\mathcal R=u/z$, where $z=\phi^\prime/H$, $\phi$ is the inflaton field, $H$ the Hubble parameter, and primes denote derivatives with respect to conformal time $\eta$ \cite{Sasaki:1986hm,Mukhanov:1988jd}.
For slow-roll inflation in the Bunch-Davies vacuum \cite{Bunch:1978yq,MFB1992}, and at first order in the slow-roll parameters $\epsilon$ and $\delta$, the coincidence limit of the two-point function formally reads \cite{baumann2012tasilecturesinflation}
\begin{equation}
\langle \mathcal R^2(\eta,\mathbf{x})\rangle
=
\int_0^\infty \frac{dk}{k}\,
P_{\mathcal R}(k,\eta),
\label{eq:2pf_intro}
\end{equation}
where 
\begin{equation}
P_{\mathcal R}(k,\eta)
=-\ \frac{ \eta k^3}{8\pi z^2}
\left|H_\nu^{(1)}(-k\eta)\right|^2
\label{PPS_class}
\end{equation}
is the dimensionless primordial curvature spectrum.  Here $\nu=3/2+2\epsilon-\delta$,   $H_\nu^{(1)}$ denotes the Hankel function of the first kind, 
\begin{equation}
    \frac1{z^2}=\frac{4\pi G}{a^2\epsilon},  
    \label{z definition}
\end{equation}
$G$ is Newton's constant, and $a$ is the scale factor.

Although the bare spectrum above is routinely employed in cosmological perturbation theory, its interpretation is problematic. The coincidence limit \eqref{eq:2pf_intro} is UV divergent and therefore requires renormalization, while cosmological observations probe finite correlation functions rather than individual Fourier amplitudes. This raises the question of whether renormalization ambiguities may induce observable corrections to the primordial spectrum. In particular, different renormalization prescriptions can modify the finite part of the two-point function and potentially affect inflationary observables such as the spectral tilt and its running \cite{del_Rio_2014}.

From the viewpoint of QFTCS, physically meaningful observables must therefore be constructed from renormalized correlation functions. This raises the central question addressed in this work: to what extent does the renormalization procedure modify the standard primordial power spectrum \eqref{PPS_class}?

In order to address this question, the first step consists in regularizing the two-point function~\eqref{eq:2pf_intro}. Its divergence structure contains both UV and infrared (IR) contributions. The UV divergence is genuine and reflects the short-distance singular structure characteristic of QFT, thus requiring regularization and renormalization. By contrast, the  IR divergence has a different physical origin: it arises from integrating over modes with arbitrarily large wavelengths, including modes whose present physical size exceeds the observable Hubble radius. Such  divergences are well understood in the literature \cite{Ferreiro_2023, Ferreiro_2024, Seery_2010} and can be controlled by introducing a  cutoff associated with an observationally inaccessible length. Since the final physical results are insensitive to the precise value of this cutoff, we omit its explicit appearance throughout the discussion for simplicity.

In cosmological spacetimes, UV-renormalization is commonly implemented through adiabatic subtraction techniques \cite{parker_toms_qft_curved_spacetime,BirrellDavies1982}. This was the original approach employed by Parker, leading to renormalized spectra significantly different from the observed one \cite{parker2007amplitudeperturbationsinflation}. Over the following decades, several variants of adiabatic renormalization were proposed, leading to different predictions for the renormalized primordial spectrum. Furthermore, the resulting physical conclusions appear to depend strongly on the particular subtraction prescription employed \cite{del_Rio_2014}.

For this reason, we adopt a conceptually different strategy. Instead of relying on adiabatic expansions, we employ the asymptotic regularization framework introduced in~\cite{Romero:2026mno}. Unlike conventional cosmological subtraction methods, asymptotic regularization isolates the UV sector directly from the asymptotic structure of the theory and connects naturally with the standard renormalization procedures used in ordinary QFT. This provides a robust and physically motivated framework to investigate the renormalization of the primordial power spectrum.


\textit{Renormalization of the primordial power spectrum.---}The UV behavior of the Bunch-Davies solutions generates the divergences appearing in the coincidence limit of the two-point function. For large momenta at fixed time ($-k\eta\to\infty$), the Hankel functions satisfy the asymptotic expansion
\begin{equation}
|H_\nu^{(1)}(-k\eta)|^2
\sim
-\frac{2}{\pi k\eta}
-\frac{\nu^2-1/4}{\pi (k\eta)^3}+\mathcal O\big((-k\eta)^{-5}\big),
\label{Hankelasymptotic}
\end{equation}
where the first two terms produce quadratic and logarithmic UV divergences, respectively.

Within the asymptotic regularization framework~\cite{Romero:2026mno}, the divergent sector is isolated directly from the UV asymptotic structure of the theory. The subtraction procedure separates the singular contributions from the finite part of the correlation function, while the remaining logarithmic sector is regularized. In this way, the UV structure is treated using the same principles that underlie standard renormalization in ordinary QFT.
This procedure also makes explicit a fundamental feature of renormalization: the presence of finite renormalization ambiguities. In general, once the UV divergent sector has been subtracted, the renormalized result is still defined up to finite contributions~\cite{Collins:105730}. Consequently, two equally valid renormalization prescriptions for the two-point function $\langle \mathcal{R}^2\rangle_{\rm ren}$ may differ by a finite contribution. Physically, this reflects the freedom in choosing a particular renormalization scheme, which fixes the finite part remaining after the divergent contributions have been removed. 

This ambiguity can be incorporated directly at the integrand level through a regular profile function $W(x)$ satisfying the   normalization condition $\mathmbox{\int d^3 \vec{x} W(|\vec{x}|)/(2\pi)^3=1}$. The renormalized two-point function can then be written as
\begin{align}
\langle \mathcal{R}^2\rangle_{\rm ren}
&=\int_0^\infty \!\!\frac{dk}{k}\Bigg[P_\mathcal{R}(k,\eta)-\frac{1}{z^2}\left(\frac{k}{2\pi}\right)^2\!\!\!\!\nonumber\\
&\ -\frac{\eta}{z^2}\left(\frac{k}{2\pi}\right)^3 \left(\pi^2\mathcal{C}W(-k\eta)-\frac{\nu^2-1/4}{(\lambda-k\eta)^3}\right)
\Bigg].
\label{renormalizedprofile}
\end{align}
Here, the terms responsible for the quadratic and logarithmic divergences in the asymptotic expansion~\eqref{Hankelasymptotic} have been subtracted following the asymptotic regularization prescription~\cite{Romero:2026mno}. In the logarithmic sector, an auxiliary  regulator $\lambda$ is introduced in order to regularize the subtraction term at the integrand level. This scale does not leave any physical imprint on the renormalized spectrum, since its dependence is absorbed into the renormalization ambiguity encoded in $\mathcal{C}$.

Expression \eqref{renormalizedprofile} naturally suggests the identification of a renormalized primordial spectrum directly from the renormalized integrand of the two-point function. However, the physically relevant regime is not the UV limit but the super-Hubble sector $-k\eta\ll1$, since observable cosmological modes lie far outside the Hubble radius by the end of inflation. 

Taking the super-Hubble limit in \eqref{renormalizedprofile}, using \eqref{z definition} and that, during slow-roll inflation, conformal time satisfies $\mathmbox{\eta=-(1+\epsilon)/(aH)}$ at first order in the slow-roll parameters, the renormalized primordial curvature spectrum becomes 
\begin{equation} P_\mathcal{R}^{\rm ren}(k,\eta)=P_\mathcal{R}(k,\eta)\Bigg[ 1-\!\left(\frac{k}{aH}\right)^{2\nu-1}\!\!\!\!\!\!\!\!\!+\widetilde {\mathcal{C}}\left(\frac{k}{aH}\right)^{2\nu}\Bigg] \label{renormalizedspectrum} \end{equation}
where the coefficient 
\begin{equation}
\widetilde {\mathcal{C}}\equiv-\frac{4\nu^2-1}{8\lambda^3}+\frac{\pi}{2}\mathcal{C}W(0)-\frac{\pi}{\nu\sin(\pi\nu)}
\end{equation}
parametrizes the finite renormalization ambiguity associated with the subtraction scheme.

Equation~\eqref{renormalizedspectrum} encapsulates the renormalization contributions relevant for the discussion that follows. The first term corresponds to the standard unrenormalized primordial spectrum, while the second term originates purely from the regularization procedure. The final contribution contains information associated with both regularization and renormalization through the finite scheme-dependent parameter $\widetilde {\mathcal{C}}$.


\textit{Robustness against renormalization ambiguities.---}We now analyze the physical impact of the renormalization ambiguity encoded in Eq.~\eqref{renormalizedspectrum}. Since all renormalization contributions appear multiplied by positive powers of $k/(aH)$, they vanish in the strict super-Hubble limit $-k\eta\to0$, where the standard inflationary spectrum is recovered. However, inflation ends before the exact limit $-k\eta\to0$ is reached. Consequently, for observable modes exiting the horizon during inflation, the renormalization contributions may still remain finite at the end of the inflationary era. In this situation, the renormalization ambiguity encoded in $\mathcal{C}$ could in principle lead to physically distinguishable predictions for the primordial spectrum.

To investigate this possibility, we focus on the most ultraviolet observable modes accessible today, corresponding to $k\simeq0.5\,{\rm Mpc}^{-1}$. Since these modes leave the Hubble radius later during inflation, they maximize the potential impact of renormalization effects and therefore provide the most conservative observational test of the framework.

For the numerical analysis presented below we set $\mathmbox{\nu=3/2}$, corresponding to the de Sitter limit. This choice is representative of slow-roll inflationary models, for which $\nu=3/2+2\epsilon-\delta$ differs from $3/2$ only by first-order slow-roll corrections \cite{MFB1992,LiddleLyth}. We have explicitly verified that varying $\nu$ within the typical slow-roll range $\nu\simeq1.51$--$1.53$ produces no appreciable changes in our results.

\begin{figure}[t]
\centering
\includegraphics[width=\columnwidth]{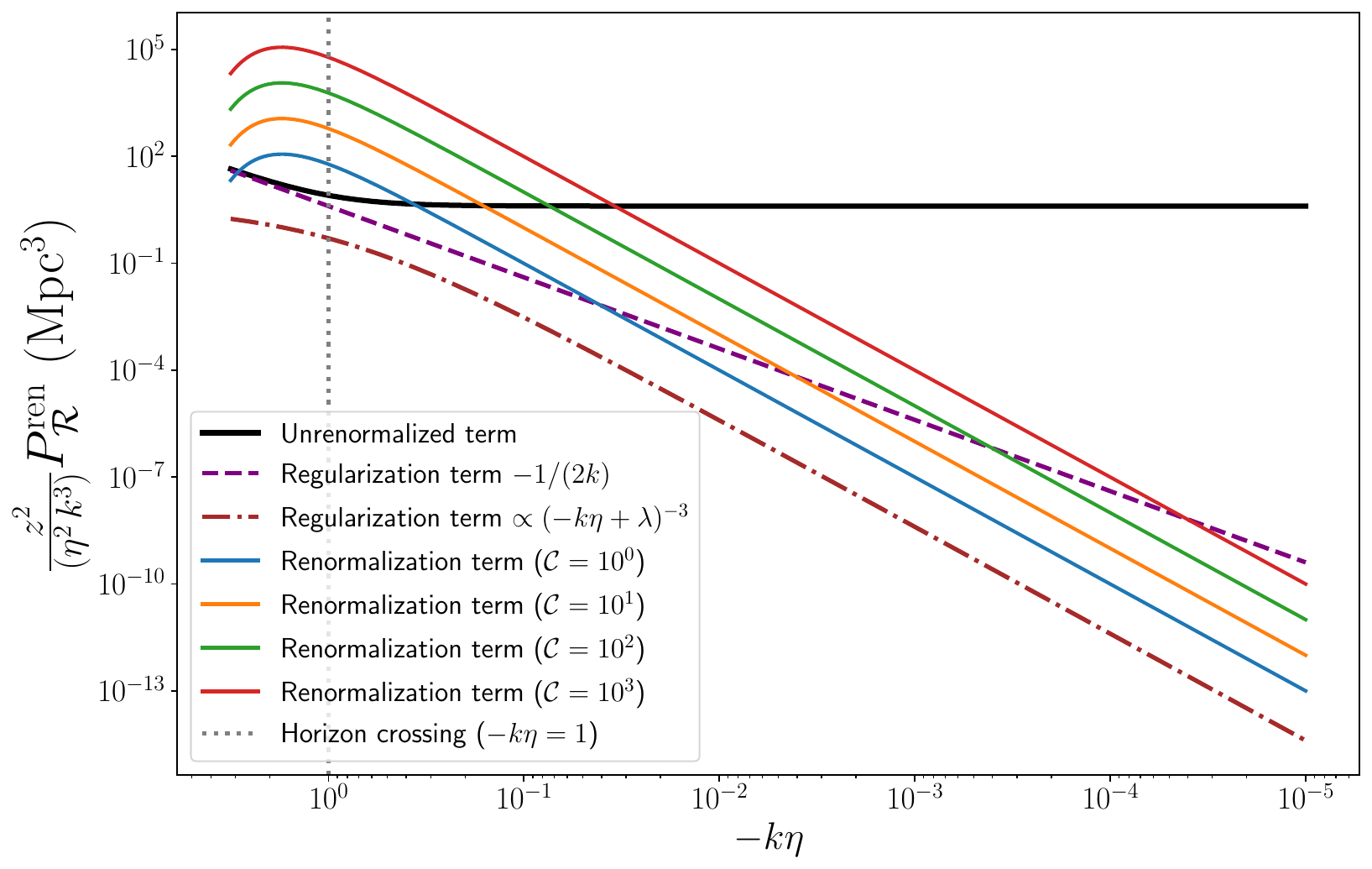}
\caption{
Individual contributions to the renormalized primordial spectrum for different renormalization schemes.
For numerical results we have chosen $W(x)=(2\pi)^{3/2}e^{-x^2/2}$ and $\lambda=1$.}
\label{fig:renormalization_terms}
\end{figure}

For several values of the renormalization parameter $\mathcal{C}$, FIG.~\ref{fig:renormalization_terms} shows the different contributions appearing in the renormalized spectrum separately, using \eqref{renormalizedprofile} without the super-Hubble approximation. The standard inflationary contribution rapidly freezes after horizon crossing, reproducing the usual super-Hubble behavior of primordial fluctuations. By contrast, the subtraction and renormalization contributions decay rapidly as inflation proceeds. This behavior reflects the fact that the renormalization sector is suppressed by positive powers of $k/(aH)$ and therefore becomes dynamically irrelevant outside the Hubble radius.

\begin{figure}[t]
\centering
\includegraphics[width=\columnwidth]{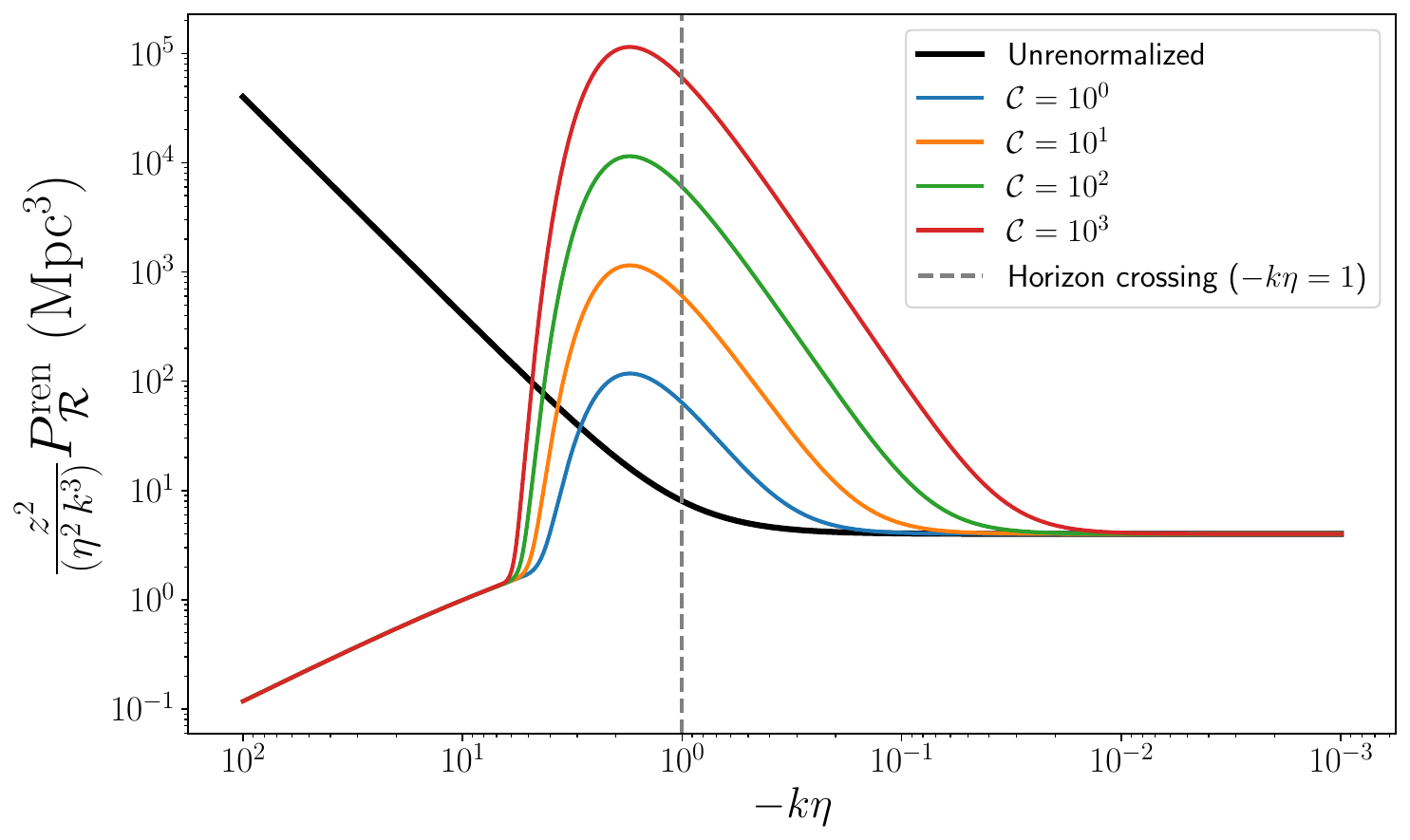}
\caption{
Comparison between the standard and renormalized primordial spectra for different renormalization schemes using $W(x)=(2\pi)^{3/2}e^{-x^2/2}$ and $\lambda=1$. On observable scales, located well beyond the region shown, all renormalized spectra converge to the standard result.}
\label{fig:renormalized_power_spectrum}
\end{figure}

This suppression mechanism explains the remarkable stability of the observable spectrum. FIG.~\ref{fig:renormalized_power_spectrum} compares the standard unrenormalized prediction with the complete renormalized spectrum for several values of $\mathcal{C}$. Even when the finite renormalization ambiguity is varied over several orders of magnitude, including values far larger than those typically expected in ordinary QFT, the renormalized spectrum rapidly converges towards the standard inflationary result. Therefore, the observable spectrum becomes progressively insensitive to the UV renormalization ambiguities encoded in $\mathcal{C}$. 

This provides a natural physical explanation for the robustness observed in FIGS.~\ref{fig:renormalization_terms} and~\ref{fig:renormalized_power_spectrum}: although different renormalization prescriptions may initially modify the spectrum, inflation dynamically drives the renormalized result towards the standard super-Hubble spectrum on observationally relevant scales.
This behavior is deeply connected with one of the key properties of inflationary perturbations: the freezing of super-Hubble modes. After horizon crossing, the amplitude of the curvature perturbation becomes approximately constant and remains conserved until horizon re-entry in the late Universe. This mechanism is what allows primordial quantum fluctuations generated during inflation to survive and eventually source the observed CMB anisotropies and large-scale structure \cite{kolbturner}.

It is important to emphasize that this conclusion does not depend on the particular choice of the profile function $W$ introduced in Eq.~\eqref{renormalizedprofile}. Since $W$ is assumed to be regular in the super-Hubble regime, its contribution only affects subleading finite parts of the renormalized spectrum and cannot modify the asymptotic  behavior relevant for cosmological observations. Consequently, different admissible choices of $W$ may change the detailed UV completion of the renormalization prescription, but they leave the observable long-wavelength physics unchanged.

To quantify the maximal observational impact of renormalization effects, we can compute the renormalized spectral index and its running,
\begin{equation}
    n_s^\mathrm{ren}-1 \equiv \frac{d\ln P^\mathrm{ren}_\mathcal{R}}{d\ln k}, \qquad \alpha^\mathrm{ren}_s \equiv \frac{dn^\mathrm{ren}_s}{d\ln k}, 
\end{equation}
using the well-established Planck 2018 constraints \cite{Planck2018}.
The running of the spectral index measures the scale dependence of the spectral tilt and therefore provides a sensitive probe of deviations from the standard inflationary prediction. In the absence of renormalization corrections, the running is highly suppressed in exact de Sitter inflation. However, the renormalized spectrum~\eqref{renormalizedspectrum} generates additional scale-dependent contributions, leading to a nontrivial running.

In order to obtain observational predictions, it is necessary to specify the inflationary evolution after horizon crossing. Observable cosmological modes correspond to perturbations that exited the Hubble radius during inflation and evolved until the end of the accelerated era, where the primordial spectrum effectively freezes.

Assuming slow-roll inflation with approximately constant Hubble parameter, horizon crossing occurs at $\mathmbox{k\simeq a_kH_k}$, while the subsequent expansion is characterized by the number of $e$-folds $N=\log a/a_k$. Using $k/(aH)\simeq e^{-N}$, the running associated with the renormalized spectrum can be written as
\begin{equation}
\alpha^\mathrm{ren}_s=-2(2\nu-1)e^{-N(2\nu-1)}\Bigg(1+\frac{3\nu}{2\nu-1}\widetilde{\mathcal{C}}e^{-N}  \Bigg).
\label{runningN}
\end{equation}
This explicitly shows the dynamical suppression of renormalization effects during inflation. Although the finite renormalization ambiguity encoded in $\widetilde {\mathcal{C}}$ may initially take large values, its observable contribution becomes exponentially diluted as inflation proceeds.

Imposing compatibility with the Planck 2018 bounds on the running of the spectral index \cite{Planck2018}, $\mathmbox{\alpha_s=-0.0045 \pm 0.0067\, (68 \% \,\mathrm{CL})}$, we derived the corresponding allowed range for the renormalization parameter $ \widetilde {\mathcal{C}}$ for inflationary scenarios with $50$ and $60$ $e$-folds. For each case, we also evaluated the relative deviation between the renormalized and standard primordial spectra,
\begin{equation}
\Delta \equiv 
\left|
\frac{P_\mathcal{R}^{\rm ren}-P_\mathcal{R}}{P_\mathcal{R}}
\right|.
\end{equation}
In order to provide a direct measure of the maximal observable impact induced by renormalization effects, we evaluate the relative error at the end of inflation using~\eqref{renormalizedspectrum}, 
\begin{equation}
    \Delta_\mathrm{end}=\Bigg[\Big( \frac{k}{aH}\Big)^{2\nu-1}\Big| 1- \widetilde{\mathcal{C}}\Big( \frac{k}{aH}\Big)\Big|\Bigg]_\mathrm{end}. 
    \label{delta_end}
\end{equation}

Equation \eqref{delta_end} provides a compact summary of the suppression mechanism discussed throughout this section. It directly quantifies the residual imprint of renormalization ambiguities on the observable primordial power spectrum at the end of inflation, after the dynamical dilution induced by the inflationary expansion has taken place. Indeed, for $N=50$, compatibility with the observational bounds allows values as large as $\mathmbox{|\widetilde{\mathcal C}|\lesssim 10^{62}}$, while the corresponding relative deviation remains below $\mathmbox{\Delta_\mathrm{end}\lesssim0.17}$. For $N=60$, the constraint becomes even weaker, $\mathmbox{|\widetilde{\mathcal C}|\lesssim 10^{76}}$, with a maximal relative deviation $\mathmbox{\Delta_\mathrm{end}\lesssim0.25}$. Finite renormalization constants associated with physically reasonable renormalization schemes are typically expected to be quantities of order unity. For such values, Eq.~\eqref{renormalizedspectrum} predicts an end-of-inflation deviation $\mathmbox{\Delta_\mathrm{end}\sim e^{-2N}}$, corresponding to $\Delta_\mathrm{end}\sim10^{-44}$ for $N=50$ and $\Delta_\mathrm{end}\sim10^{-52}$ for $N=60$. Observable corrections become appreciable only when $\widetilde{\mathcal C}$ is taken to be extraordinarily large, of order $10^{62}$--$10^{76}$, far beyond what would normally be regarded as physically reasonable.

\begin{figure}[t]
\centering
\includegraphics[width=\columnwidth]{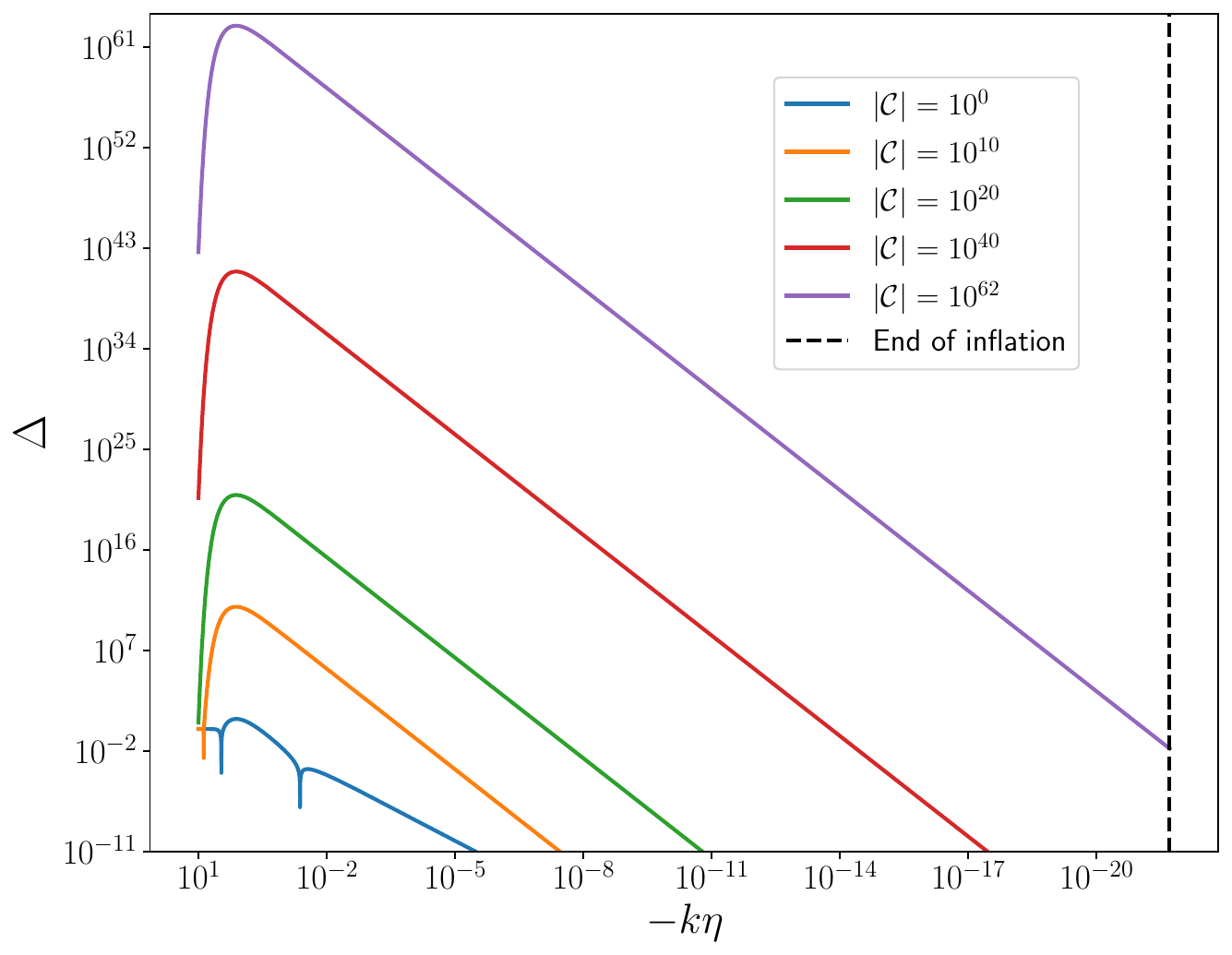}
\caption{
Relative deviation between the renormalized and standard primordial spectra using different renormalization constants for the relevant scales $k\simeq0.5\,{\rm Mpc}^{-1}$.}
\label{fig:relative_error}
\end{figure}

FIG.~\ref{fig:relative_error} illustrates the evolution of the relative deviation $\Delta$ between the renormalized and standard primordial spectra from horizon crossing up to the end of inflation for several representative large values of the renormalization parameter $\mathcal C$ in the $N=50$ $e$-fold scenario. Near horizon crossing, renormalization effects can dominate the spectrum, leading to large relative deviations. However, as the mode evolves into the super-Hubble regime, these corrections, even for extremely large values of the renormalization ambiguity, decrease rapidly due to the inflationary suppression encoded in Eq.~\eqref{renormalizedspectrum}. In particular, values as large as $|\mathcal C|\sim1$--$10^{40}$ already lead to relative deviations below $\Delta\sim10^{-11}$ well before the end of inflation. Consequently, observable effects survive only for extraordinarily large values of $\mathcal C$, close to the conservative upper bounds derived from observational constraints.

\textit{Discussion.---}
We have studied the renormalization of the scalar primordial power spectrum using the asymptotic framework developed in Ref.~\cite{Romero:2026mno}. Within this approach, the subtraction sector and the finite renormalization ambiguities both decay outside the Hubble radius as positive powers of $k/(aH)$. As a consequence, inflation dynamically suppresses the renormalization sector while the physical super-Hubble perturbations freeze, rendering the observable spectrum progressively insensitive to UV ambiguities.

This behavior differs conceptually from previous analyses based on adiabatic subtraction prescriptions~\cite{silviapla}, where a scale-invariant subtraction contribution survives throughout inflation and only becomes suppressed after the transition to a post-inflationary FRW phase. In the present framework, by contrast, the potentially relevant renormalization contributions are already diluted during inflation itself, so that the renormalized spectrum approaches the standard frozen super-Hubble result directly within the inflationary era.

We have also quantified the observational impact of finite renormalization ambiguities by constraining an effective renormalization parameter through current bounds on the running of the spectral index. Even allowing values many orders of magnitude larger than those typically expected in ordinary QFT, the resulting corrections to the observable spectrum remain remarkably small. Our analysis therefore leads to the conclusion that the large renormalization ambiguities previously discussed in the literature do not translate into significant observable effects within physically admissible renormalization schemes.

The present analysis has been carried out for scalar perturbations. However, the underlying suppression mechanism is independent of the spin of the field, and renormalization effects on the primordial tensor spectrum are therefore similarly diluted during inflation.

\textit{Conclusions.---}
The primordial power spectrum is ultimately a quantity derived from a renormalized quantum two-point function. Therefore, the relevant question is not whether renormalization modifies the spectrum, but whether physically admissible renormalization prescriptions lead to observable consequences.

Our results indicate that they do not. Although renormalization ambiguities are unavoidable at the formal level, inflation dynamically suppresses their contribution to observable quantities. As modes evolve outside the Hubble radius, the renormalization sector is exponentially diluted while the physical perturbations freeze, making the primordial spectrum extraordinarily insensitive to UV details. The apparent tension between renormalization in QFTCS and the standard inflationary predictions therefore seems largely illusory: inflation itself protects the observable spectrum from UV ambiguities.

\textit{Acknowledgments.---}
We are very grateful to Antonio López Maroto and Javier Rubio Peña for their valuable guidance and insightful discussions on the interpretation, formal aspects, and applicability of the results presented in this work.
This work was financially supported by Grant PID2023-149018NB-C44 funded by
MCIN/AEI/10.13039/501100011033 and by ``ERDF/EU''.
RN was funded by the Royal Society through the University Research Fellowship Renewal URF$\backslash$R$\backslash$221005. CDR acknowledges financial support from
MCIU (Ministerio de Ciencia, Innovación y Universidades, Spain) fellowship FPU24/03217.

\bibliographystyle{apsrev4-2}
\bibliography{biblio}

\end{document}